\documentstyle[twocolumn,aps]{revtex}

\begin{document}
\draft
\title{Shor-Preskill type security-proof for the quantum key
 distribution without public announcement of bases}

\author{Won-Young Hwang $^1$ \cite{email}
        Xiang-Bin Wang $^1$ , Keiji Matsumoto $^1$,
       Jaewan Kim $^2$, and Hai-Woong Lee $^3$} 

\address{$^1$ IMAI Quantum Computation and Information Project,
ERATO, Japan Science and Technology Corporation,
Daini Hongo White Bldg. 201, 5-28-3, Hongo, Bunkyo,
Tokyo 133-0033, Japan}

\address{$^2$ School of Computational Sciences,
Korea Institute for Advanced Study, Seoul 130-012, Korea }    

\address{$^3$ Department of Physics,
Korea Advanced Institute of Science and Technology, Daejeon   
305-701, Korea }
\maketitle
\begin{abstract}
We give a Shor-Preskill type security-proof
to the quantum key distribution without public
 announcement
of bases [W.Y. Hwang {\it et al.}, Phys. Lett. A {\bf 244},
 489 (1998)].
 First, we modify the Lo-Chau protocol once more so that it
 finally reduces to the quantum key distribution without public
 announcement of bases.
Then we show how we can estimate the
error rate in the code bits based on that in the checked bits in
the proposed protocol, that is the central point of the
proof. We discuss the problem of imperfect sources and that of
large deviation in the error rate distributions.
We discuss when the bases sequence must be discarded.
\end{abstract}
\section{introduction}
Information processing with quantum systems enables what
seems to be impossible with its classical counterpart
\cite{wies,shor,dowl,delg}.
In addition to the practical importance, this fact has many
theoretical and even philosophical implications \cite{lloy,deut}.

Quantum key distribution (QKD)
\cite{bene,eker,ben2,ben3,brus,gisi}
is one of the most important and interesting
quantum information processing.
QKD will become the first practical quantum information
 processor \cite{gisi}.
 Although the security of the Bennett-Brassard 1984 (BB84)
QKD scheme
\cite{bene} had been widely conjectured based on the no-cloning
theorem \cite{woot,diek},
it is quite recently that its unconditional security was
shown \cite{maye,biha,sho2}.
In particular, Shor and Preskill \cite{sho2}  showed the
security of BB84 scheme, starting from a modified form of
 the Lo-Chau protocol \cite{lo3}, by elegantly using the
 connections among several basic
ideas in quantum information processings, e.g. quantum error
correcting codes (QECCs) \cite{cald,stea} and entanglement
purification \cite{ben4}.

In the standard BB84 protocol, however, only half of the data
 obtained by using expensive quantum communication can be
 utilized at most.  It is clear that it is not efficiency but
 security that is the most important in the cryptographic tasks.
 However, it is meaningful enough to improve the efficiency
without loss of security.
One method for the full efficiency QKD is to delay the
 measurements in the BB84 scheme using quantum memories.
This is indeed the original proposal by Bennett and Brassard
 \cite{bene}.
However, the quantum memories would be quite costly with
 near-future technology. Another method is to assign
 significantly different probabilities to the different bases
\cite{lo2}. Although unconditional security of the scheme is
given \cite{lo2}, it has a disadvantage that a larger number of
 key must be generated at once than in the BB84 scheme in
order to get the same level of security. However, in a
 recently proposed QKD without public announcement of basis
 (PAB) \cite{hwa2}, we can obtain the full efficiency with such
 problem relaxed.

The QKD without PAB is a simple variation of BB84 scheme. In the
 BB84 scheme, Alice and Bob use different random bases and
then discard the cases where the bases are not matched. In the
 QKD without PAB, Alice and Bob use bases determined by
 a prior random key, the basis sequence $b$.
When the basis sequence $b$ is used only once, it is clear that
 the scheme is as secure as the BB84 scheme. However, in this
 case it is obviously meaningless because they have to consume
 secret key that is  as long as the generated key. Thus, the
 problem is that whether the basis sequence $b$ can be 
 repeatedly used without loss of security. It was shown that it is the
 case against the individual attacks \cite{hwa2} and it was 
 suggested that it could be against the
 coherent attacks \cite{hwan}.
The purpose of this paper is to give the Shor-Preskill type
unconditional security proof to the QKD without PAB.
The framework of the proof is the same as the
original one \cite{sho2}. However, we modify the Lo-Chau
 scheme once more to give the QKD without PAB. We give
 three schemes: modified Lo-Chau scheme II that
 reduces to Calderbank-Shor-Steane (CSS) codes
 scheme II \cite{cald,stea}. The CSS codes scheme II then
 reduces to the QKD without PAB.
We argue why we can estimate the
error rate in the code bits based on that in the checked bits in
the protocol, that is the central point of the
proof. This implies that the
modified Lo-Chau scheme II is secure, completing the proof.
We discuss the problem of imperfect sources and that of
large deviation in the error rate distributions.
We discuss when the bases sequence must be discarded.
Then we give a conclusion.
\subsection{Notation}
In this paper,  we use mostly the notations in Refs. \cite{sho2,lo}.

 The canonical basis of a qubit consists of
$|0\rangle$ and $|1\rangle$. We define another basis as follows.
$|\bar{0}\rangle= (1/\sqrt{2})(|0\rangle+|1\rangle)$ and
$|\bar{1}\rangle= (1/\sqrt{2})(|0\rangle-|1\rangle)$.
The Hadamard transform $H$ is a single qubit unitary
transformation of the form
$H= (1/\sqrt{2}) 
( \begin{array}{cc} 1 & 1  \\
                                 1 & -1
             \end{array}
)$ in the canonical basis.
This transformation interchanges the bases
$|0\rangle$, $|1\rangle$ and $|\bar{0}\rangle$, $|\bar{1}\rangle$.
 $I= \sigma_0$ is the identity operator and
$\sigma_x= ( \begin{array}{cc} 0 & 1  \\
                                       1 & 0
             \end{array} ),
 \sigma_y= ( \begin{array}{cc} 0 & -i  \\
                                            i &  0
             \end{array} ),
\sigma_z= ( \begin{array}{cc} 1 & 0  \\
                                         0 & -1
             \end{array} )$
are the Pauli operators.
The $\sigma_{{a}(i)}$ denotes the Pauli operator $\sigma_{a}$
acting on the $i$-th qubit where $a=0,x,y,z$.
For a binary vector $s$, we let
$\sigma_a^{[r]}=
\sigma_{{a}(1)}^{s_1} \sigma_{{a}(2)}^{s_2}
 \cdot \cdot \cdot \sigma_{{a}(n)}^{s_n} $,
 where $s_i$ is the $i$-th bit of $s$ and
$\sigma_a^0= I$, $\sigma_a^1=\sigma_a $.

The Bell basis states are the four maximally entangled ones, 
$|\Psi^{\pm}\rangle =(1/\sqrt{2})(|01\rangle \pm |10\rangle)$ and
$|\Phi^{\pm}\rangle =(1/\sqrt{2})(|00\rangle \pm |11\rangle)$.

Let us consider two classical binary codes, $C_1$ and $C_2$,
such that $\{0\} \subset C_2 \subset C_1 \subset F_2^n $ where
$ F_2^n $ is the binary vector space of the $n$ bits.
A set of basis for the CSS code can be obtained from vectors
$v \in C_1$ as follows,
$v \rightarrow (1/|C_2|^{1/2}) \sum_{w \in C_2}|v+w \rangle $.
Note that $v_1$ and $v_2$ give the same vector if
$v_1-v_2 \in C_2$.
$H_1$ is the parity check matrix for the code $C_1$ and
$H_2$ is that for $C_2^\perp$, the dual of $C_2$.
 $Q_{x,z}$ is a class of QECCs. For $v \in C_1$, the corresponding
 code word is
 $v \rightarrow (1/|C_2|^{1/2}) \sum_{w \in C_2}
  (-1)^{z \cdot w}|x+v+w \rangle$.
\section{the QKD without public announcement of basis}
 It is notable that what
 we are considering in this section is not security but
 reductions of the schemes.

 {\it Protocol A: Modified Lo-Chau scheme II.}\\
(1) Alice creates $2n$ Einstein-Podolsky-Rosen (EPR) pairs in the
state   
$|\Phi^+\rangle ^{\otimes 2n}$. 
(2) Alice and Bob are assumed to be sharing a prior random
$(2n/r)$-bit string, the basis sequence $b$. 
(2n/r is a positive integer.)
Alice performs the Hadamard transform on second half of each
EPR pair for which $b$ is one. 
(3) Alice repeats the step 2 $r$ times with the same basis
 sequence $b$. 
(4) Alice sends the second half of each pair to Bob. 
(5) Bob receives the qubits and publically announces this fact.
(6) Bob performs the Hadamard
 transform on second half of each EPR pair for which 
 $b$ is one. 
(7) Bob repeats the step 6 $r$ times with the same basis
 sequence $b$.
(8) Alice randomly selects $n$ of the $2n$ EPR pairs to serve
as check bits to test for Eve's interference. Then
she announces it to Bob. 
(9) Alice and Bob each measure their halves of the $n$ check EPR
pairs in the $ \{ |0\rangle,|1\rangle \} $ basis and share the
results. If too many of these measurements disagree, they abort
the scheme. 
(10) Alice and Bob make the measurements
on their code qubits of
$\sigma_z^{[r]}$ for each row $r \in H_1$ and
$\sigma_x^{[r]}$ for each row $r \in H_2$. Alice and Bob share
the results, compute the syndromes for bit and phase flips,
and then transforms their state so as to obtain $m$
(encoded) nearly EPR pairs. 
(11) Alice and Bob measure the EPR pairs in
 the (encoded) $\{|0\rangle, |1\rangle\}$
basis to obtain $m$-bit final string with near-perfect security.
$\Box$

The entanglement purification   
 protocols with one-way classical communcations are equivalent to
 the QECCs \cite{ben4}.
The modified Lo-Chau protocol
 reduces to the CSS codes protocol by this equivalence
 \cite{sho2}. However, the
only difference between the Protocol A and the modified Lo-Chau
 protocol is the following. In the former they use the basis
 sequence $b$ to determine whether they apply the Hadamard
operation or not, while in the latter
they do it by their own different random sequences and they
use only matched bases.
We can see that the protocol $A$ reduces to the
protocol $B$ by the same equivalence.

 {\it Protocol B: CSS codes scheme II.}\\
(1) Alice creates $n$ random check bits
and a random $m$-bit key $k$. They are assumed to share
 a prior random $(2n/r)$-bit string, the basis sequence $b$. 
(2) Alice chooses $n$-bit strings $x$ and $z$
at random. 
(3) Alice encodes her key $|k\rangle$ using the CSS code
$Q_{x,z}$. 
(4) Alice chooses $n$ positions out of $2n$
and puts the check bits in these positions and the code bits
in the remaining positions. 
(6) Alice performs the Hadamard transform on the qubits
for which $b$ is one. 
(7) Alice repeats the step 6 $r$ times with the same basis
sequence $b$. 
(8) Alice sends the resulting state to Bob. Bob acknowledges the
receipt of the qubits. 
(9) Alice announces the positions of check bits, the values
 of the check bits, $x$, and $z$. 
(10) Bob performs the Hadamard transform on  
 the qubits  for which the component of $b$ is one. 
(11) Bob repeats the step 10 $r$ times with the same basis
sequence $b$. 
(12) Bob checks whether too many of the
 check bits have been corrupted, and aborts the scheme if so.
(13) Bob measures the qubits in the
(encoded) $\{|0\rangle, |1\rangle\}$ basis to obtain
 $m$-bit final key with near-perfect security.
 $\Box$

The only difference between the Protocol $B$ and the CSS codes
 protocol \cite{sho2}
 is the following. In the former they use the basis
 sequence $b$ to determine whether they apply the Hadamard
operation or not, while in the latter
they do it by their own different random sequences and they
use only matched cases.
 We can see
that the protocol $B$ reduces to the following protocol $C$ in
the same way as the modified CSS codes protocol reduces to
 the BB84 protocol.

{\it Protocol C: QKD without public announcement of basis}\\
(1) Alice creates $2n$ random bits. 
 Alice and Bob are sharing a prior random $(2n/r)$-bit string,
the basis sequence $b$.
(2) Alice encodes each random bit to qubits
 using the basis sequence $b$.
That is, when the random bit is $0$ ($1$) and
 the corresponding component of the basis sequence
$b$ is zero, she creates a qubit in the
$ |0\rangle$ ($|1\rangle$) state.
When the random bit is $0$ ($1$) and
 the corresponding component of the basis sequence
$b$ is one, she creates a qubit in the
$ |\bar{0}\rangle$ ($|\bar{1}\rangle$) state. 
(3) Alice repeats the step 2 $r$ times with the same basis
sequence $b$. 
(4) Alice sends the resulting qubits to Bob. 
(5) Bob receives the $2n$ qubits and performs measurement
 $S_z$ or $S_x$
if the corresponding component of the sequence
$b$ is zero and one, respectively.
Here $S_z$ ($S_x$) is the
orthogonal measurements whose eigenvectors are
 $|0\rangle $ and $|1 \rangle$
($|\bar{0}\rangle$ and $|\bar{1} \rangle$).
(6) Bob repeats the step 5 $r$ times with the same basis 
 sequence $b$.
(7) Alice decides randomly on a set of $n$ bits
 to use for the protocol. Then she announces it.
The other $n$ qubits are used as check-bits. 
(8) Alice and Bob announce the values of the
 their check-bits. If too few of these values agree, they abort the
 protocol. 
(9) Alice announces $u+v$, where $v$ is a string consisting of
randomly chosen code-bits,
and $u$ is a random code word in $C_1$. 
(10)  Bob subtracts
$u+v$ from his  code-bits, $v+\epsilon$, and
corrects the result, $u+\epsilon$, to a codeword in $C_1$. 
(11)  Alice and Bob use the coset of each $u+C_2$ as the key. In
this way, they obtain $m$-bit string. $\Box$
\section{the security of the QKD without PAB}
Since we have shown the reduction of protocols
 $A \rightarrow B \rightarrow C$, it is sufficient for us to show
 the security of the protocol $A$ here.
 Arguments in the following are for
entanglement purifications in the protocol $A$, thanks to which
we can deal with the coherent attacks.

We briefly remind the classicalization of statistics that is
stressed by Lo and Chau \cite{lo3,lo}. Then we will see that
remaining arguements are similar to what we used for the
individual attacks \cite{hwa2}.

First, let us review the classicalization of
statistics in the Shor and Preskill proof \cite{sho2}. What we
consider is
the interaction of qubits $|\psi\rangle$ of
 Alice and Bob and
quantum probes $|e\rangle$ of Eve.
 In general, the state after any
 interaction by a unitary operator $U$ can be decomposed
\cite{sho3,pres} as
\begin{equation}
\label{a}
U|\psi\rangle |e\rangle=
\sum_{\{k\}} C_{\{k\}}
\sigma_{{k_1}(1)} \sigma_{{k_2}(2)} \cdot \cdot \cdot
 \sigma_{{k_n}(n)}
|\psi\rangle |e_{\{k\}} \rangle.
\end{equation}
Here $\{k\}$ is the abbreviation for the $k_1,k_2,...,k_n$ with
 $k_i= 0,1,2,3$ ($i=1,2,...,n$), and $\sigma_0= I$,
 $\sigma_1= \sigma_x$, $\sigma_2= \sigma_y$,
 $\sigma_3= \sigma_z$.
The $C_{\{k\}}$'s are coefficients.
The vectors $|e_{\{k\}} \rangle$ are nomalized but not mutually
orthogonal in general.
Since Eq. (\ref{a}) is just the geneal decomposition of
 a vector by complete bases, it is clear that the
 interaction described in Eq. (\ref{a}) includes the case of the
 coherent attacks as well as individual attacks.
It is notable that Eve can make her quantum probes interact with
 Alice and Bob's qubits only when she has access to their
 qubits. In other words, Eve cannot modifty the interaction
 after the qubits left her. This is in contrast with the fact
 that Eve can
choose the measurement bases even after the qubits left.
Therefore we need not worry about Eve's later choice
if our consideration is for the interaction term Eq. (\ref{a}).
What Eve can do is only to control the coefficients
$C_{\{k\}}$'s as she likes.

Let us note that the each state
 $\sigma_{{k_1}(1)} \sigma_{{k_2}(2)}
\cdot \cdot \cdot \sigma_{{k_n}(n)} |\psi \rangle$
is an eigenstate of the measurements that are performed here.
The qubits are initially prepared in the state 
 $|\Phi^+\rangle$ that is one of the Bell states.
 The set of the Bell states are closed for Pauli operations on
a qubit. Thus each qubit in the protocol that has undergone
a certain Pauli operation is one of the Bell states. On the other
hand, the measurements performed in the checking steps is
equivalent to the Bell measurements \cite{sho2}.
Therefore, as long as the checking measurements are concerned,
the state in a mixed state
\begin{eqnarray}
\label{b}
\rho &=&
\sum_{\{k\}} |C_{\{k\}}|^2
\sigma_{{k_1}(1)} \sigma_{{k_2}(2)}
 \cdot \cdot \cdot \sigma_{{k_n}(n)} |\psi\rangle
\langle \psi| \sigma_{{k_1}(1)} \sigma_{{k_2}(2)}
\nonumber\\
    && \cdot \cdot \cdot \sigma_{{k_n}(n)},
\end{eqnarray}
gives rise to the same results as the pure state in Eq. (\ref{a}).
This is the basis for the classicalization of statistics \cite{lo3,lo},
as a result of which it is sufficient for us to consider
classical distributions given by probabilities
 $P_{\{k\}}= |C_{\{k\}}|^2$ .

Once the classicalization of statistics is
obtained, it is not difficult to see that the modified Lo-Chau
 protocol II is secure.
In the case of the BB84 protocol, they estimate the error rate, or
the ratio of
 $\sigma$'s that are not identity operator
 $I$ among the $\sigma_{{k_1}(1)} \sigma_{{k_2}(2)}
 \cdot \cdot \cdot \sigma_{{k_n}(n)}$'s, by doing the checking
 measurement on some randomly chosen subsets of the qubits.
 If Eve's operation on a checked qubit is the identity $I$, the
 probability to give rise to error is zero. If Eve's operation on
a checked qubit is not the identity $I$, it will give rise to errors
 probabilistically: If the basis matches it will induce no error but
 if the bases do not match it will.
(More precisely, the probabilities to give rise to
 errors is $1/2$, $1/2$,
 and $1$, respectively, for $\sigma_z$, $\sigma_x$, $\sigma_y$
 operations.)
 What Eve wants to do is to minimize the number of
 errors in the check bits for a given number of non-identity
 operations. However, since the checked bits and the bases are
 randomly chosen by Alice and Bob, Eve knows nothing about
 them while she has access to the qubits of Alice and Bob.
Thus we can assume that the
 error rate of the checked bits represents that of the code bits,
 that is a crucial point in the security proof.

Let us now consider the Protocol A. The Protocol A to the first
 round is obviously stronger than the modified Lo-Chau protocol.
Thus it is clear that to the first round the Protocol A is as secure
 as the modified Lo-Chau protocol. Let us consider the
 second round. Here one may worry about that Eve can extract
some information about the basis sequence $b$ after the first
 round. It is obvious that if Eve knows the basis
 sequence $b$ she can successfully cheat. It is because in
 this case she can control the probabilities $P_{\{k\}}$'s so that
 more bases are matched or the probability to be detected
 decreases. However, no matter how many rounds are performed
 Eve can extract no information on the basis sequence $b$ by any
 quantum operations in the ideal case \cite{hwa2}: The ensemble
 of qubits with different bases
give rise to the same density operators. (We will discuss the
 non-ideal case in the next section. Also note that all public
discussions between Alice and Bob are performed after all qubits
 have arrived at Bob in the proposed protocol.) So we don't have
 to worry about this point.
Now what Eve knows is that the same basis sequence $b$ is used
 repeatedly. That is, she knows which and which qubits are in the
 same basis although she does not know the identity of the basis.
 {\it Now the problem is that whether Eve can induce statistically
 smaller number of errors in the checked bits for a given number
of non-identity operators in the second round than in the first
 round. However, we can see that she cannot do so because
 she does not know which basis it is anyway and thus the
 probabiltity that the basis are not matched is still 1/2.} For
 example, let us consider the first two qubits in the first and
second round. If Eve's basis and the basis of checked bit is
 matched (not matched) then the probability that it is to be
detected is zero (non-zero). Even if Eve knows that the two
qubits are in the same basis, that information is not helpful in
decreasing the expected error rate since the probabiltity that
the basis are not matched is still 1/2. Eve's best strategy here is
 to choose the same operations for the two qubits. Then
 although the average error rate is not changed, the
 deviation of the probabilistic distribution will be increased. (We
 will discuss about the problem of the large deviation in the next
section.) We can easily see that the same argument applies to
remaining qubits and all qubits in the later $j$-th rounds
($j=3,4,5,...,r$).
Therefore we can safely estimate the error rate in the code bits
based on that in the checked bits, as we did in the modified
 Lo-Chau protocol \cite{sho2}.
\section{discussion and conclusion}
Let us consider the problem of the imperfect sources.  As noted
 in the previous section, the following fact is crucial for the QKD 
without PAB. 
 The two ensemble of states,  that is, the equal mixture of the
 $|0\rangle$ and $|1\rangle$ and that of the $|\bar{0} \rangle$
 and $|\bar{1} \rangle$ are equivalent to each other and thus
 cannot be distinguished in any case. This is valid when the
sources are ideal.
However, there must be a certain amount of imperfection in the
source. In this case some amount of
 information on the basis sequence $b$ can be leaked to Eve,
 making  the scheme insecure \cite{refe}. However, we give
 a practical method to overcome this problem.
 It is not difficult to generate pairs of qubits in one of the
 (imperfect) Bell state, for example, the  $|\Phi^+\rangle$ state,
with current technologies \cite{gisi}.
Alice can generate the qubits to be sent to Bob
 in the following way.  First she
 prepares pairs of qubits in the (imperfect)
 $|\Phi^+\rangle$ state and
 she performs either the measurement $S_z$ or $S_x$
on one qubit of each pair. Here $S_z$ ($S_x$) is the
orthogonal measurements whose eigenvectors are
 $|0\rangle $ and $|1 \rangle$
($|\bar{0}\rangle$ and $|\bar{1} \rangle$).
She sends the other unmeasured qubits to Bob.
Bob's ensemble of qubits generated by $S_z$ ($S_x$) is a mixture
of imperfect $|0\rangle $ or $|1 \rangle$
(either $|\bar{0}\rangle$ or $|\bar{1} \rangle$).
However, these two ensembles cannot be distinguished
in principle.
It is because Alice's different choice of measurement cannot
change the density operator of Bob's ensemble.
Thus at least the problem of leakage of
 the information about the basis sequence $b$ can be overcome.
However, this does not mean that the QKD without PAB with
 imperfect source is secure. This problem is beyond the scope
of this paper.
The Shor-Preskill paper \cite{sho2} shows the
security with perfect sources only.
The security with imperfect source has been dealt with
recently \cite{inam}.

Next, let us compare the efficient QKD \cite{lo2} with the QKD
without PAB. In the former, they obtain the efficiency
 $\epsilon^2+ (1-\epsilon)^2$ for a given $0 < \epsilon \leq 1/2$.
 The number of check bits in the other basis is proportional
 to $\epsilon^2$. Thus, when $\epsilon$ is small, namely when the
 efficiency is nearly full, the former would have the problem
 of small number of samples for data analysis.
In order to obtain enough security, therefore, they have to
distribute a large number of qubits at once.
 In the latter we have
 a similar problem in a different way, as we noted in the previous
 section. That is, if Eve had chosen the same operation for the
qubits with the same bases, the deviation in the probabilistic
distribution of the error rate of the checked bits would be larger
 than that of the BB84 protocol, for a given number of total data,
 $n$. However, the random sampling process to estimate
 the error rate in the first round with $n/r$ bits will be at least as
 good as that of the BB84 protocol with the same $n/r$ bits.
 That is, the error rate
deviation of the QKD without PAB with $r$ rounds of $n/r$ bits will
 be at least as small as that of the BB84 protocol with $n/r$ bits.
 (We can see that the former is strictly smaller than the latter.)
 Therefore, provided that the length $n/r$ of the basis sequence is
 long enough we can say that the proposed protocol is secure.

If the error rate in the checked bits is too high
because of noise on the communication line or because of Eve,
the protocol is aborted.
One may worry about that some information about the basis
sequence has leaked to Eve in this case.
If Alice use again the random bits to be encoded
 (in step (1) of the Protocol C)
with the same basis sequence $b$, it amounts to that
the qubits in the same state are repeatedly used. Then it is
simple for Eve to get information about the basis sequence.
However, if the random bits are newly generated
everytime, the two ensembles of qubits corresponding to different
bases have the same density operator $I$ and thus they cannot
be distinguished, as we have discussed. Therefore, as long as
Alice uses the ramdom bits to be encoded only once,
they don't have to discard the basis sequence $b$ even after
the protocol had been aborted because of high error rate.

However,
it should also be underlined \cite{hwa2} that the basis sequence
has to be discarded after the final key is used for encrypting
a message, because a ciphertext gives partial information about
 the key by which it is encrypted. The information about the key
can then used to extract information about the basis sequence
 $b$.

In conclusion, we have given a Shor-Preskill type
security-proof
to the quantum key distribution scheme without public
 announcement
of basis \cite{hwa2}. We have given the modified Lo-Chau   
 protocol II. This scheme reduces to the CSS codes scheme II
 that reduces to the QKD without PAB. We have reviewed how
 the classicality is obtained in the Shor-Preskill type proof. Using
 the classicality we argued how we can estimate the error
rate in the code bits based on that in the checked bits in the
 modified Lo-Chau protocol II.
Since remaining arguments are the
same, this completes the proof. We discussed the problem of
 imperfect source and that of necessity of generation of a large
 number of data.
We discussed when the bases sequence must be discarded.
\acknowledgments
We are very grateful to Dr. Yuki Tokunaga in NTT for helpful
comments. W.-Y. H., X.-B. W., and K. M. are
very grateful to Japan Science Technology Corporation
for financial supports.
H.-W. L. appreciate the financial support from the Brain Korea 21
Project of Korean Ministry of Education. J.K. was supported by
Korea Research Foundation Grant 070-C00029.    


\begin{references}
\bibitem[*]{email}  Present address: Department of Electrical and 
Computer Engineering,
Northwestern University, 
Evanston, IL 60208, USA;
Email address: wyhwang@ece.northwestern.edu


\bibitem{wies} S. Wiesner, Sigact News {\bf15}(1), 78  (1983).
\bibitem{shor} P. Shor, Proc. 35th Ann. Symp. on Found. of
               Computer Science. (IEEE Comp. Soc. Press,
               Los Alomitos, CA, 1994) 124-134.
\bibitem{dowl} J.P. Dowling, Phys. Rev. A {\bf 57}, 4736 (1998).
\bibitem{delg} A. Delgado, W.P. Schleich, and G. S\"{u}ssmann,
                New Journal of Physics {\bf 4}, 37.1 (2002).
\bibitem{lloy} S. Lloyd, Complexity 3(1), 32 (1997),
               quant-ph/9912088
               (available at http://xxx.lanl.gov).
\bibitem{deut} D. Deutsch, A. Ekert, and R. Lupacchini,
                       math-HO/9911150.
\bibitem{bene} C.H. Bennett and G. Brassard, in : Proc. IEEE
                Int. Conf. on Computers, systems, and signal
                processing, Bangalore (IEEE, New York, 1984)
                p.175.
\bibitem{eker} A.K. Ekert, Phys. Rev. Lett. {\bf67}, 661 (1991).
\bibitem{ben2} C.H. Bennett, G. Brassard, and N.D. Mermin, Phys.
               Rev.Lett. {\bf68},  557 (1992).
\bibitem{ben3} C.H. Bennett,
               Phys. Rev. Lett. {\bf68}, 3121  (1992) ;
               A.K. Ekert, Nature {\bf358}, 14  (1992).
\bibitem{brus} D. Bru\ss, Phys. Rev. Lett. {\bf81},
               3018  (1998).
\bibitem{gisi} N. Gisin, G. Ribordy, W. Tittel, H. Zbinden,
               Rev. Mod. Phys. {\bf 74}, 145 (2002),
               references therein.
\bibitem{woot} W.K. Wootters and W.H. Zurek, Nature  {\bf299},
                802  (1982).
\bibitem{diek} D. Dieks, Phys. Lett. A {\bf92}, 271 (1982).
\bibitem{maye} D. Mayers,
                 J. Assoc. Comput. Mach. {\bf 48}, 351 (2001).
\bibitem{biha} E. Biham, M. Boyer, P.O. Boykin, T. Mor, and
               V. Roychowdhury, in {\it Proceedings of the
               Thirty-Second Annual ACM Symposium on Theory
               of Computing} (ACM Press, New York, 2000),
               pp.715-724, quant-ph/9912053.
\bibitem{sho2} P.W. Shor and J. Preskill, Phys. Rev. Lett.
               {\bf 85}, 441 (2000).
\bibitem{lo3}  H.-K. Lo and H.F. Chau, Science {\bf283},
              2050 (1999).
\bibitem{cald} A.R. Calderbank and P.W. Shor, Phys. Rev. A
               {\bf54}, 1098  (1996).
\bibitem{stea} A.M. Steane, Phys. Rev. Lett. {\bf77},
               793 (1996).
\bibitem{ben4} C.H. Bennett, D.P. DiVincenzo, J.A. Smolin, and
               W.K. Wootters, Phys. Rev. A {\bf54},
               3824  (1996).
\bibitem{lo2} H.-K. Lo, H.F. Chau, and M. Ardehali,
quant-ph/0011056.
\bibitem{hwa2} W.Y. Hwang, I.G. Koh, and Y.D. Han,
                 Phys. Lett. A   {\bf 244}, 489 (1998)
\bibitem{hwan} W.Y. Hwang, D. Ahn, and S.W. Hwang,
               Phys. Lett. A {\bf 279}, 133 (2001).
\bibitem{lo} H.-K. Lo, Quan. Inf. Com. {\bf 1}, 81 (2001).
\bibitem{sho3} P. Shor, Phys. Rev. A {\bf 52}, 2493 (1995).
\bibitem{pres} J. Preskill,
            Proc. Roy. Soc. Lond. A {\bf 454}, 385 (1998).
\bibitem{refe} An anonymous referee of one of the author's
 other  previous work pointed out this, to whom we are grateful.
\bibitem{inam} H. Inamori, N. L\"{u}tkenhaus, D. Mayers,
                  quant-ph/0107017.
\end{references}
\end{document}